\documentclass[pra,aps,a4paper,twocolumn]{revtex4}
\usepackage{amsfonts}
\usepackage{graphicx}
\usepackage{amsmath}

\newcommand\mX{{\mathcal X}}
\newcommand\mR{{\mathcal R}}
\newcommand\mZ{{\mathcal Z}}
\newcommand\mG{{\mathcal G}}
\newcommand\mI{{\mathcal I}}
\newcommand\mS{{\mathcal S}}
\newcommand\mQ{{\mathcal Q}}
\newcommand\bs{{\boldsymbol s}}
\newcommand\bt{{\boldsymbol t}}
\newcommand\bc{{\boldsymbol c}}
\newcommand\bb{{\boldsymbol b}}
\newcommand\ba{{\boldsymbol a}}
\newcommand\be{{\boldsymbol e}}

\begin{document}

\title{Graphical Nonbinary Quantum Error-Correcting Codes}
\author{Dan Hu, Weidong Tang, Meisheng Zhao, and Qing Chen}
\affiliation{$^1$Hefei National Laboratory for Physical Sciences at
Microscale and Department of Modern Physics, University of Science
and Technology of China, Hefei, Anhui 230026, China}
\author{Sixia Yu$^{1,2}$ and C.H. Oh}
\affiliation{$^2$Physics Department, National University of
Singapore, 2 Science Drive 3, Singapore 117542}
\date{\today}

\begin{abstract}
In this paper, based on the nonbinary graph state, we present a
systematic way of constructing good non-binary quantum codes, both
additive and nonadditive, for systems with integer dimensions. With
the help of computer search, which results in many interesting codes
including some nonadditive codes meeting the Singleton bounds, we
are able to construct explicitly  four families of optimal codes,
namely, $[[6,2,3]]_p$, $[[7,3,3]]_p$, $[[8,2,4]]_p$ and
$[[8,4,3]]_p$  for any odd dimension $p$ and a family of nonadditive
code $((5,p,3))_p$ for arbitrary $p>3$. In the case of composite
numbers as dimensions, we also construct a family of stabilizer
codes $((6,2\cdot p^2,3))_{2p}$ for odd $p$, whose coding subspace
is {\em not} of a dimension that is a power of the dimension of the
physical subsystem.
\end{abstract}

\maketitle

\section{Introduction}

Noises are inevitable and they cause errors in quantum informational
processes. One active way of dealing with errors is provided by the
quantum error-correcting codes (QECCs)
\cite{qecc1,qecc2,qecc3,qecc4}, which have found many applications
in quantum computations and quantum communications, such as the
fault-tolerant quantum computation \cite{ftc}, the quantum key
distributions \cite{qkd}, and the entanglement purification
\cite{ep,gtm}. Roughly speaking, a QECC is a subspace of the Hilbert
space of a system of many physical subsystems with the property that
the quantum data encoded in this subspace can be recovered
faithfully, even though a certain number of physical subsystems may
suffer arbitrary errors, by suitable syndrome measurements followed
by corresponding unitary transformations.

An important family of QECCs is the {\it stabilizer code}
\cite{gott1,crss,calder}, which is specified by the joint +1
eigenspace of a stabilizer, an Abelian group of tensor products of
Pauli operators. The stabilizer formalism has also been established
in the nonbinary case \cite{E.Knill,nonbinary,nonbinary1,MarKus,Ket} and
many good codes for systems of a prime or a power of prime dimension
have been constructed \cite{MarKus,Ket,chau,feng}, including the
well-known perfect code with five registers \cite{chau} for all
dimensions. Though the majority of QECCs constructed so far are
stabilizer codes, including the CSS codes \cite{css}, the
topological codes \cite{tc1}, color codes \cite{tc2}, and also the
recently introduced entanglement-assisted codes \cite{eaqecc}, there
are a few exceptions called as {\it nonadditive codes}
\cite{rshor,rains,smol,9123}.

The nonadditive code does not admit a stabilizer structure and it
should not be a subcode of some larger stabilizer code with the same
distance otherwise it will be a trivial nonadditive code. Though
difficult to construct and identify the nonadditive codes promise a
larger coding subspace since less structured than the stabilizer
codes. For qubits the nonadditive error-correcting code that
outperforms the stabilizer codes has been constructed \cite{9123}
based on the binary graph states. A graphical approach \cite{SiXia},
as well as a codeword stabilized code approach \cite{zb}, to the
construction of binary additive and nonadditive codes has been
developed based on the binary graph states.

The graph states \cite{werner,graph} are useful multipartite
entangled states that are essential resources for the one-way
computing \cite{oneway} and can be experimentally demonstrated
\cite{six}. The binary graph state proves to an extremely effective
tool \cite{werner,grsl,SiXia} in the construction of QECCs. The
nonbinary graph states were introduced first in \cite{werner} and
discussed in details in the case of systems of an odd prime
dimension \cite{Mohsen}. It is also investigated in the context of
universal quantum computation \cite{duanl} and other applications
\cite{nonbinary2}. Recently an approach to construct QECCs based on
nonbinary graph states has been introduced in \cite{looi} and some
new codes are found via computer search for qubits and qutrits.

Here we shall generalize the graphical construction of QECCs
\cite{SiXia} to the nonbinary case based on the nonbinary graph
states from which some analytic constructions are attainable. In
Sec.II the nonbinary graph states  are introduced.  In Sec.III we
introduce the concept of coding clique for a weighted graph and show
how it is related to the construction of both stabilizer and
nonadditive QECCs.  In Sec.IV we present some codes found via
numerical searches and the graphical versions of some known codes as
illustrations. In Sec.V we construct analytically four families of
optimal stabilizer codes that saturate the quantum Singleton bound
for any odd dimension as well as a family of nonadditive codes
$((5,p,3))_p$ for all $p>3$. In Sec.VI we investigate the graphical
codes arising from composite systems and construct a family of
stabilizer codes $((6,2\cdot p^2,3))_{2p}$ with $p$ being odd. The
graph states of systems composed of coprimed subsystems are in a
one-to-one correspondence with the direct product of the graph
states of subsystems.

\section{Nonbinary graph states}

Here we shall consider a general system with $p$ levels, a {\it
qupit} for short, where $p$ is arbitrary. We denote by $\mathbb
Z_p=\{0,1\ldots, p-1\}$ the ring with addition modulo $p$. Under the
computational basis $\{|i\rangle|i\in \mathbb Z_p\}$ of a qupit, the
generalized bit shift and phase shift operators
 read
\begin{equation}
\mX=\sum_{l\in\mathbb Z_p}|l+1\rangle\langle l|,\quad
\mZ=\sum_{l\in\mathbb Z_p}\omega^{l}|l\rangle\langle
l|,\quad\left(\omega=e^{ i\frac {2\pi}p}\right).
\end{equation}
Obviously $\mX^p=\mZ^p=\mI$ and $\mZ\mX=\omega\mX\mZ$. Here we denote by $\bar \mX$
the Hermitian conjugate of $\mX$. The
computational basis $|l\rangle$ is the eigenstate of $\mZ$ with
eigenvalue $\omega^l$ while the eigenstate of $\mX$ with eigenvalue
$\omega^j$ reads
\begin{equation}
|\theta_j\rangle=\frac1{\sqrt p}\sum_{l\in \mathbb
Z_p}\omega^{-lj}|l\rangle.
\end{equation}

A $\mathbb Z_p$-weighted graph $G=(V,\Gamma)$ is composed of a set
$V$ of $n$ vertices and a set of weighted edges specified by the
{\it adjacency matrix} $\Gamma\in \mathbb Z_p^{n\times n} $, an
$n\times n$ matrix with zero diagonal entries and the matrix element
$\Gamma_{ab}\in\mathbb Z_p$ denoting the weight of the edge
connecting vertices $a$ and $b$. The graph state associated with a
given weighted graph $G=(V,\Gamma)$ of a system of $n$ qupits
labeled with $V$ reads \cite{werner}
\begin{equation}
|\Gamma\rangle=\frac1{\sqrt{p^n}}\sum_{\boldsymbol s\in \mathbb
Z_p^V}\omega^{\frac12\boldsymbol s\cdot \Gamma\cdot\boldsymbol
s}|\boldsymbol s\rangle=\prod_{a,b\in V}\mathcal ({\mathcal
U}_{ab})^{\Gamma_{ab}}|\theta_0\rangle^V.
\end{equation}
Here we have denoted by $\mathbb Z^V_p$ the set of all the vectors
$\bs=(s_1,s_2,\ldots,s_n)$ with $n$ components $s_a\in \mathbb Z_p$
$(a\in V)$, by $|\boldsymbol s\rangle$ the common eigentsate of all
the phase shifts $\mZ_a$ $(a\in V)$ with eigenvalue $\omega^{s_a}$,
by
\begin{equation}
|\theta_0\rangle^V=|\theta_0\rangle_{1}\otimes|\theta_0\rangle_{2}\otimes\ldots
|\theta_0\rangle_{n} \end{equation} the joint $+1$ eigenstate of all
bit shifts $\mX_a$ $(a\in V)$, and by
\begin{equation}
\mathcal {U}_{ab}=\sum_{i,j\in\mathbb
Z_p}\omega^{ij}|i\rangle\langle i|_a\otimes |j\rangle\langle j|_b
\end{equation}
the non-binary controlled phase gate between two qupits $a$ and $b$.
The non-binary graph state $|\Gamma\rangle$ is also the unique (up
to a global phase factor) joint $+1$ eigenstate of the following $n$
vertex stabilizers
\begin{equation}
\mG_a=\mX_a\prod_{b\in V}(\mZ_b)^{\Gamma_{ab}}, \quad a\in V.
\label{hu}
\end{equation}

For ${\bs}\in\mathbb Z_p^V$ we denote
$\mX^{\bs}=\mX_1^{s_1}\mX_2^{s_2}\ldots\mX_n^{s_n}$ and similarly
for the phase shift operator $\mZ^{\boldsymbol s}$. Obviously
\begin{equation}
\mG_\bs\equiv\prod_{a\in
V}(\mG_a)^{s_a}=\omega^{\frac12\bs\cdot\Gamma\cdot\bs}\mX^\bs\mZ^{\bs\cdot\Gamma}
\end{equation}
is also a stabilizer of the graph state for arbitrary $\bs\in
\mathbb Z_p^V$, i.e., $\mG_\bs|\Gamma\rangle=|\Gamma\rangle$. All
the stabilizers of the graph state belong to the generalized Pauli
group for qupits
\begin{equation}\label{pg}
P_n=\{e^{-i\frac{\pi}{p}(p-1){\bs}
\cdot{\bt}}\mX^{\bs}\mZ^{\bt}| {\bs, \bt} \in
\mathbb{Z}_{p}^V\}\times \{\omega^l|l\in \mathbb Z_p\}.
\end{equation} The {\it
graph-state basis} of the $n$-qupit Hilbert space $\mathcal {H}_n$
refers to $\{|\Gamma_{\bc}\rangle \equiv
\mZ^{\bc}|\Gamma\rangle|{\bc}\in \mathbb Z_p^V\}$.  Under the computational
basis a graph-state basis looks like
\begin{equation}
|\Gamma_\bc\rangle=\mZ^\bc|\Gamma\rangle=\frac1{\sqrt{p^n}}\sum_{\boldsymbol
s\in \mathbb Z_p^V}\omega^{\frac12\boldsymbol s\cdot
\Gamma\cdot\boldsymbol s+\bc\cdot\bs}|\boldsymbol s\rangle.
\end{equation}
A collection of
$K$ different vectors $\{{\bc}_1,{\bc}_2,\ldots,{\bc}_K\}$ in
$\mathbb Z_p^V$ specifies a $K$ dimensional subspace of $\mathcal
{H}_n$ that is spanned by $K$ graph-state basis
$\{|\Gamma_{{\bc}_i}\rangle\}_{i=1}^K$.

\begin{figure}
\includegraphics{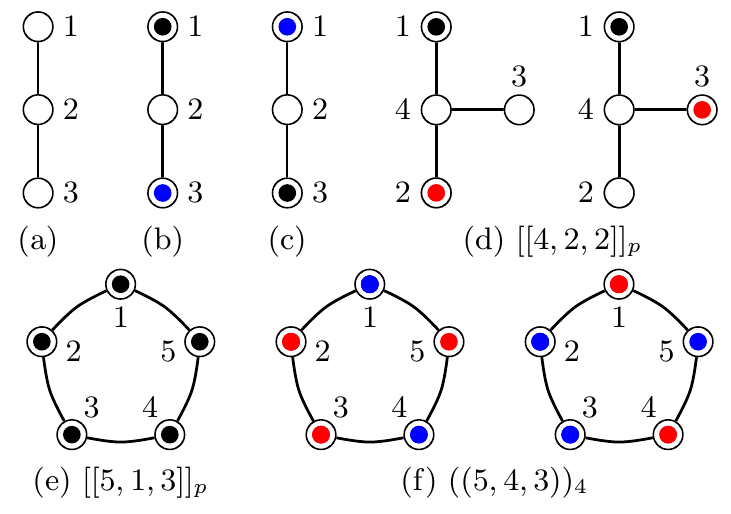} \caption{\label{tu1} Graph, graph states, and
graph-state basis. (a-c) Star graph on 3 qubits and graph-state
basis; (d) Star graph on 4 vertices; (d-e) Loop graph on 5 vertices.
Colored vertices represent also a vector in $\mathbb Z_p$: white,
black, blue, and red vertices having weights $0,1,2,p-1$
respectively. Black edges have weight 1.}
\end{figure}

For an example the graph state corresponding to the star graph $S_3$
on 3 vertices with all edges weighted 1 as shown in Fig.\ref{tu1}(a)
represents the GHZ state
\begin{equation}
|S_3\rangle=\frac1{\sqrt p}\sum_{j\in\mathbb
Z_p}|\theta_{p-j}\rangle_1\otimes|j\rangle_2\otimes|\theta_{p-j}\rangle_3.
\end{equation}
An edge with weight 1 will be represented by a black line and an
edge with weight $p-1$ will be represented by a red thick line as in
Fig.\ref{tu2}. In addition we will also indicate a vector in
$\mathbb Z_p^V$ via colored vertices with white, black, blue, and
red vertices representing weights $0,1,2,p-1$ respectively. For
example in the graph shown in Fig.\ref{tu1}(b) a vector
$(1,0,2)\in\mathbb Z_p^3$ is indicated via the colored vertices and
therefore we have a graph-state base
\begin{equation}
\mZ_1\mZ_3^2|S_3\rangle=\frac1{\sqrt p}\sum_{j\in\mathbb
Z_p}|\theta_{p-j-1}\rangle_1\otimes|j\rangle_2\otimes|\theta_{p-j-2}\rangle_3.
\end{equation}

\section{Coding cliques and QECCs}

From the standard theory of the QECC we know that if a set of Pauli
errors can be corrected then all the errors can also be corrected.
The situation in non-binary case is the same, because the nonbinary
error basis defined by $\mathcal {E}=\{\mX^{\bs}\mZ^{\bt}| {\bs,
\bt} \in \mathbb{Z}_{p}^V\}$ is a nice basis \cite{E.Knill, Ket},
i.e., the set $\mathcal {E}$ forms a basis for the operators acting
on the $n$-qupit Hilbert space $\mathcal {H}_n$.

For a given weighted graph $G=(V,\Gamma)$, from the definition the
vertex stabilizers of a graph state, we have
\begin{equation}
\mX^{\bs}\mZ^{\bt}=\omega^{-\frac12\bs\cdot\Gamma\cdot\bs}\mG_\bs\mZ^{\bt-\bs\cdot\Gamma}.
\end{equation}
That is to say every non-binary Pauli error acting on the graph
state $|\Gamma\rangle$ can be equivalently replaced by some qupit
phase flip errors, up to some phase factors. For convenience we
refer that the vector $\bt-\bs\cdot \Gamma$ is {\it covered} by the
error $\mX^\bs\mZ^\bt$. Given an integer $d$ we introduce a {\it
$d$-uncoverable set} as
\begin{equation}
\mathbb D_d=\mathbb
Z_p^V-\left\{{\bt}-{\bs}\cdot\Gamma\Big|0<|\sup({\bs})\cup\sup({\bt})|<d\right\},
\end{equation}
where we have denoted by $\sup({\bf s})=\{a\in V|s_a\neq 0\}$ the
support of a vector ${\bs}\in \mathbb Z_p^V$ and by $|C|$ the number
of the elements in $C\subseteq V$.  That is to say $\mathbb D_d$ is
the set of all the $d$-uncoverable vectors, i.e., vectors cannot be
covered by Pauli errors acting nontrivially on less then $d$ qupits,
in $\mathbb Z_p^V$. In addition we define the {\it $d$-purity set}
as
\begin{equation}
\mathbb S_d=\left\{{\bs}\in \mathbb
Z_p^V\Big||\sup({\bs})\cup\sup({\bs}\cdot\Gamma)|<d\right\}.
\end{equation}
It is obvious that if $\bs \in \mathbb S_d$  then the graph
stabilizer $\mG_\bs$  has a support less than $d$, i.e., act
nontrivially on less than $d$ qupits.

A {\it coding clique} $\mathbb{C}^K_d$ of a given graph
$G=(V,\Gamma)$ is a collection of $K$ different vectors in $\mathbb
Z_p^V$ that satisfy:
\begin{itemize}
\item[i)]  ${\bf 0} \in\mathbb{C}^{K}_d$;
\item[ii)]  ${\bs}\cdot{\bc}=0$ for all ${\bs}\in\mathbb S_d$ and
every ${\bc}\in \mathbb C^K_d$;
\item[iii)]  ${\bc}-{\bc}^\prime\in \mathbb D_d$ for all ${\bc}, {\bc}^\prime\in \mathbb C^K_d$.
\end{itemize}
If the coding clique $\mathbb{C}^K_d$  forms a group with respect to
the addition modulo $p$, then it will be referred to as a {\it
coding group}. In words, a coding clique $\mathbb C^K_d$ is a
collection of $K$ $d$-uncoverable vectors in $\mathbb Z_p^V$ that is
orthogonal to the vectors in $\mathbb S_d$ and the difference
between any two vectors in $\mathbb C_d^K$ is also $d$-uncoverable.
We can build a super graph $\mathbb G$ with vertices being the
vectors in $\mathbb D_d\cap\mathbb S_d^\bot$, i.e., $d$-uncoverable
vectors that are orthogonal to all the vectors in $\mathbb S_d$, and
two vertices in the super graph are connected by an edge iff their
difference is also $d$-uncoverable. Then the coding clique $\mathbb
C_d^K$ is exactly a $K$-clique, a subset of the vertices which are
pairwise connected, of the super graph $\mathbb G$. This fact
justifies the nomenclature {\it clique}.

\textbf{Theorem 1} Given a graph $G=(V,\Gamma)$ and one of its
coding clique $\mathbb C_d^K$, the subspace $(G, K, d)_p$ spanned by
the graph-state basis $ \left\{|\Gamma_{{\bc}}\rangle|{\bc}\in
\mathbb C^K_d\right\}$ is an $((n, K, d ))_p$ code, which is a
stabilizer code if $\mathbb C_d^K$ is a group and a nonadditive code
if $\mathbb C_d^K$ is neither a group nor a subset of a coding group
$\mathbb C_{d^\prime}^{K^\prime}$ of $G$ with $d^\prime\ge d$.

{\bf Proof.} To prove that the subspace spanned by the graph-state
basis  $\{|\Gamma_{{\bc}}\rangle|{\bc}\in \mathbb C^K_d\}$ is an
$((n, K, d ))_p$ code we have only to show that the condition
\cite{qecc3,werner}
\begin{equation}\label{cond}
\langle\Gamma_{\bc}|\mathcal
{E}_{d}|\Gamma_{\bc^\prime}\rangle=f(\mathcal
{E}_{d})\delta_{\bc\bc^\prime}
\end{equation}
is fulfilled for all $\bc, \bc^\prime\in\mathbb{C}^K_d$ and all the
error $\mathcal {E}_{d}=\mX^{\bs}\mZ^{\bt}$ that acts nontrivially
on a number  of qupits that is less than $d$, i.e.,
$|\sup({\bs})\cup\sup({\bt})|<d.$

If the error is proportional to a stabilizer of the graph state,
i.e, $\mathcal E_d=f(\bs)\mG_\bs$ for some $\bs\in \mathbb Z_p^V$
and phase $f(\bs)$, then $\bs\in \mathbb S_d$. Because of condition
ii) of the coding clique the error acts like a constant operator on
the subspace $(G,K,d)$ so that the condition Eq.(\ref{cond}) is
fulfilled with a $f(\mathcal {E}_{d})=f(\bs)$ that is independent of
$\bc$. If the error $\mathcal {E}_{d}$ is neither one of the
stabilizers of the graph state $|\Gamma\rangle$ nor the identity
operator, i.e., $ \mathcal E_d=\mX^{\bs}\mZ^{\bt}\propto
\mG_\bs\mZ^{\bt-\bs\cdot\Gamma} $ with $\bt\neq\bs\cdot\Gamma$, then
$\langle\Gamma_\bc|\mathcal
{E}_{d}|\Gamma_{\bc^\prime}\rangle\propto \langle\Gamma|\mathcal
{Z}^{\be}|\Gamma\rangle$ with
$\be=\bc^\prime-\bc+\bt-\bs\cdot\Gamma$. By virtue of condition iii
of the coding clique it is ensured that $\be\neq \boldsymbol 0$ for
all $\bc,\bc^\prime \in \mathbb C^K_d$ which gives rise to
Eq.(\ref{cond}) with $f(\mathcal E_d)=0$. Thus we have proved that
$(G, K, d)_p$ is an $((n, K, d ))_p$ code.

By definition, a nonbinary stabilizer code is the joint +1
eigenspace of an Abelian subgroup of generalized Pauli group $P_n$
given in Eq.(\ref{pg}). The subspace $(G,K,d)_p$ spanned by the
graph-state basis $\{|\Gamma_{{\bc}}\rangle|{\bc}\in \mathbb
C^K_d\}$ is stabilized by an element $\mS$ of $P_n$ if and only if
$\mS$ is one of the stabilizers of the graph state $\mG_\bs$  with
$\bs\in \mathbb S$ where
\begin{equation}\label{cs}
\mathbb S=\left\{\bs\in\mathbb Z_p^V\big|\bs\cdot \bc=0, \forall\
\bc\in\mathbb C_d^K\right\}.
\end{equation}

If $\mathbb C_d^K$ is a coding group then it is obviously an Abelian
group and can be generated by a set of vectors $\langle
\bc_1,\bc_2,\ldots,\bc_k\rangle$ in $\mathbb Z_p^V$
 with degrees $[\mu_1,\mu_2,\ldots,\mu_k]$, i.e., the minimal
 positive number such that $\mu_i\bc_i=0$ (mod $p$) $(i=1,2,\cdots k)$.
It is easy to see that all $\mu_i$'s divide $p$ so that
$K=\prod_{i=1}^k\mu_i$ divides $p^n$. Furthermore
$(\omega=e^{i2\pi/p})$
\begin{equation}
p^n=\sum_{\bc\in \mathbb C^K_d}\sum_{\bs\in \mathbb
Z_p^V}\omega^{\bs\cdot \bc}=\sum_{\bs\in \mathbb Z_p^V}\sum_{\bc\in
\mathbb C^K_d}\omega^{\bs\cdot \bc}=|\mathbb S|K
\end{equation}
because ${\bf 0}\in \mathbb C_d^K$ for the first equality and
\begin{equation}
\sum_{\bc\in \mathbb C^K_d}\omega^{\bs\cdot
\bc}=\prod_{i=1}^k\sum_{m_i=0}^{\mu_i-1}\omega^{m_i\bs\cdot
\bc_i}=\left\{\begin{matrix}K &\mbox{if}\quad \bs\in \mathbb
S,\\\\
0&\mbox{if}\quad \bs\not\in \mathbb S,\end{matrix}\right.
\end{equation}
(since $\mu_i\bc_i=0$ (mod $p$)) for the last equality. That is to
say if $\mathbb C_d^K$ is a group then we can find exactly a number
$p^n/K$ of stabilizers $\{\mG_\bs|\bs\in \mathbb S\}$ whose joint +1
eigenspace is exactly $(G,K,d)_p$. Thus the code $(G,K,d)_p$ is a
stabilizer code.

If $\mathbb C_d^K$ is not a group then we denote by $\mathbb C$ the
group generated by $\mathbb C_d^K$, i.e., the smallest group that
contains $\mathbb C_d^K$. Obviously the subspace $\mQ$ spanned by
$\{|\Gamma_\bc\rangle|\bc \in \mathbb C\} $ is the joint +1
eigenspace of stabilizer $\{\mG_\bs|\bs\in \mathbb S\}$ and
$|\mathbb S|<p^n/K$. If the subspace $\mQ$ can detect $d-1$ or more
errors then $\mathbb C$ is a coding group of $G$. If the subspace
$\mQ$ cannot detect $d-1$ errors then any subset of the stabilizer
cannot either. On the other hand every stabilizer code that contains
$(G,K,d)_p$ must have a stabilizer that is a subset of the
stabilizer of $\mQ$. Therefore if $\mathbb C_d^K$ is not a subset of
some coding group $\mathbb C_{d^\prime}^{K^\prime}$ of $G$ with
$d^\prime\ge d$ then the code $(G,K,d)$ is a nonadditive code.
 \hfill Q.E.D.

Similar conclusions about the stabilizer codes appeared also in
\cite{looi}. If all the generators of a coding group have the
maximal degree $p$ then corresponding stabilizer code can be denoted
as $[[n,k,d]]_p$. However there are cases where the generators of
the coding group are not all of maximal degree. Then we have still a
stabilizer code but the dimension of the code subspace may {\em not}
be a power of $p$ and we shall denote such a stabilizer code as
$((n,\mu_1\cdot\mu_2\cdot \ldots\cdot\mu_k,d))_p$. A family of such
kind of stabilizer codes will be provided in Sec. VI.

\section{graphical Nonbinary  QECCs via numerical search}

According to Theorem 1 we can use the same systematic algorithm
developed for binary case in \cite{SiXia} to do a systematic search
for the non-binary quantum codes, i.e.
\begin{itemize}
\item[i)] To input a $\mathbb Z_p$-weighted graph $G=(V,\Gamma)$ on $n$ vertices;
\item[ii)] To choose a distance $d$ and compute the $d$-purity
set $\mathbb S_d$ and the $d$-uncoverable set $\mathbb D_d$ so that
a super $\mathbb{G}$ can be built;
\item[iii)] To find all the
$K$-clique $\mathbb{C}^K_d$ of the super graph $\mathbb{G}$;
\item[iv)] To output a $(G, K, d)_p$, i.e., an $((n, K, d ))_p$ code that is spanned by the
basis $\{|\Gamma_{\bc}\rangle|{\bc}\in \mathbb C_d^K\}$.
\end{itemize}

In practice, we have used the clique finding program {\it cliquer}
\cite{Niskanen} to search for the cliques for the super graph.
Within our present computation power systematic search for graphical
codes can be done up to $p=6$ and $n=6$ and some tentative searches
have been done for $n=8$. In what follows a quantum code will be
specified by a weighted graph together with a coding clique or the
generators of a coding group. Though the search for cliques are
hard, the verifications of them are relative easy.


\subsection{The code $[[3,1,2]]_3$}
The first example is the stabilizer code $[[3,1,2]]_3$, which is known and
has been constructed, e.g., in \cite{MarKus}. We consider the
star graph $S_3$ on 3 vertices with two edges all weighted 1 as shown in Fig.\ref{tu1}(a).
A coding group generated by
$(1,0,2)\in\mathbb Z_3^3$ as shown in Fig.\ref{tu1}(b) provides the code $[[3,1,2]]_3$, i.e., a
3-dimensional subspace spanned by the graph-state basis
\begin{equation}
\big\{\mZ_1^a\mZ_3^{2a}|S_3\rangle | a\in \mathbb Z_3 \big\}.
\end{equation}
The stabilizer of this code is generated by $\langle \mG_2,
\mG_1\mG_3\rangle$ with $\mG_a$ being defined in (\ref{hu}). It is
easy to see that the stabilizer is equivalent to the stabilizer $\langle \mX_{123},
\mZ_{123}\rangle$ appeared in \cite{gtm} under a local unitary transformation.
It is not difficult to see that we can have a $[[3,1,2]]_p$ for any odd $p$
with the same graph and the same coding group generated by
$(1,0,-1)\in\mathbb Z_p^3$ with stabilizer $\langle\mG_2,\mG_1\mG_3\rangle$.

\subsection{The code $((3,3,2))_4$}
For an odd number of qupits with $p$ being even there is no code of
distance 2 that saturates the Singleton bound so far. We have found
a suboptimal code $((3,3,2))_4$ instead. For the equal weighted star
graph on 3 vertices as shown in Fig.\ref{tu1}(a) we have  found a
coding clique: $$\{ (0,0,0), (1,0,2), (2,0,1)\},$$  with
corresponding graph-state basis shown in Fig.\ref{tu1}(a-c), which
span the code subspace of a nonadditive $((3,3,2))_4$ code. In fact,
as we see later, we can construct a code $((3,p-1,2))_p$ with even
$p$.

\subsection{The code $[[4,2,2]]_6$}
The next example we concerned is also a 1-error detecting code
$[[4,2,2]]_6$ which can be constructed from the star graph $S_4$ on 4
vertices as shown in Fig.\ref{tu1}(d). From this graph a 2 dimensional coding
group can be found to be generated by vectors $(1,-1,0,0)$ and
$(1,0,-1,0)$ in $\mathbb Z_6^4$. That is to say the code
$[[4,2,2]]_6$ is the 36-dimensional subspace spanned by the basis
\begin{equation}
\left\{\mZ_1^{-a-b}\mZ_2^a\mZ_3^b|S_4\rangle\ \big|\
a,b\in \mathbb Z_6\right\}.
\end{equation}
whose stabilizer is generated by $\langle \mG_1\mG_2\mG_3,\mG_4\rangle$.
On the same graph we also find another coding clique as
\begin{equation}
\left\{(a+b,-a,-b,\delta_{a1}\delta_{b1})\in \mathbb Z_6^4\big|
a,b\in \mathbb Z_6\right\}
\end{equation}
which can be stabilized by $\langle \mG_1\mG_2\mG_3\rangle$ only
thus we have a nonadditive $((4,36,2))_6$ code. The nonadditive
codes meeting the Singlet Bound is a very common situations in the
nonbinary graphical codes, almost every stabilizer codes will have a
clique set which are not a group, which means we can always
construct a nonadditive code from a graphical nonbinary  stabilizer
code.

\begin{figure}
\includegraphics{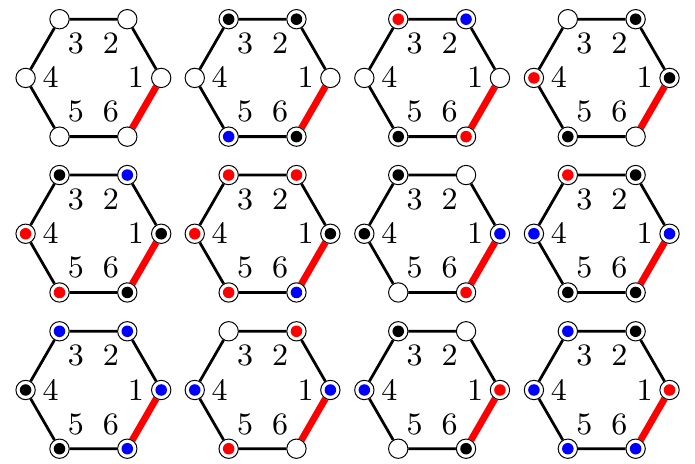} \caption{\label{tu2} Twelve graph-state bases for
the code $((6,12,3))_4$. The thick red edge has a weight $p-1$.}
\end{figure}

\subsection{The code $[[5,1,3]]_3$}

The first example of 1-error-correcting code is the well-known
$[[5,1,3]]_3$ code, which is the only error-correcting codes
for arbitrary dimension so far and was constructed in
\cite{chau} and formulated with graph state in \cite{werner}. For the loop
graph $L_5$ on 5 vertices with all edges weighted 1
we find a 1-dimensional
coding group generated by $(1,1,1,1,1)\in \mathbb Z_3^5$
as indicated in Fig.\ref{tu1}(e). It is clear that this coding group
can be generalized to arbitrary dimension and we obtain for arbitrary $p$ the
$[[5,1,3]]_p$ code that is spanned by graph-state basis
\begin{equation}
\left\{\mZ_1^a\mZ_2^a\mZ_3^{a}\mZ_4^{a}\mZ_5^{a}|L_5\rangle\ \big|\
a\in \mathbb Z_p\right\}.
\end{equation}

\subsection{The code $((5,4,3))_4$}

For dimension 4 we also found a nonadditive code $((5,4,3))_4$ that
saturates the quantum Singleton bound. The graph we considered is still
the
loop graph $L_5$ on 5 vertices with all edges weighted 1. The coding
clique contains the following 4 vectors in $\mathbb Z_4^5$
\begin{equation*}
\big\{(0,0,0,0,0),(1,1,1,1,1), (2,3,3,2,3), (3,2,2,3,2)\big\}
\end{equation*}
among which one basis is shown in Fig.\ref{tu1}(e) and two bases are
shown in Fig.\ref{tu1}(f). Since there is a stabilizer of the code
containing $4^3<4^4$ elements only and it satuates the Singleton
bound, the code is not a subcode of any 1-error correcting code and
is therefore nonadditive. In fact we can construct a nonadditive
$((5,p,3))_p$ for any $p> 3$ as will shown later.

\subsection{The code $((6,12,3))_4$}

We also found a nonadditive code $((6,12,3))_4$ on the loop graph
$L_6$ on 6 vertices with one edge weighted $3$ represented by a
thick red edge as in Fig.\ref{tu2}.  The coding clique $\mathbb
C_3^{12}$ of $L_6$ with 12 vectors reads
\[ \begin{matrix}
\begin{matrix}
 &$(0 0 0 0 0 0)$ &$(0 1 1 0 2 1)$ &$(0 2 3 0 1 3)$\\
 &$(1 1 0 3 1 0)$ &$(1 2 1 3 3 1)$ &$(1 3 3 3 2 3)$\\
 &$(2 0 3 1 0 3)$ &$(2 1 3 2 1 1)$ &$(2 2 2 1 1 2)$\\
 &$(2 3 0 2 3 0)$ &$(3 0 1 2 0 1)$ &$(3 1 2 2 2 2)$\\
\end{matrix}
\end{matrix} \]
which is represented in Fig.\ref{tu2} with white, black, blue, and
red vertices having weights 0,1,2,3 respectively. There are only 4
stabilizers can be found for this code, namely
$$\{\mI,\mG_3^2\mG_6^2,(\mG_1\mG_2\mG_4\mG_5)^2\mG_3\mG_6,(\mG_1\mG_2\mG_4\mG_5)^2\bar\mG_3\bar\mG_6\},$$
whose joint +1 eigenspace  cannot be any 1-error correcting code
because of the Singleton bound. Therefore the code is nonadditive.

\subsection{The codes $[[7,3,3]]_3$ and $[[8,4,3]]_3$}

\begin{figure}
\includegraphics{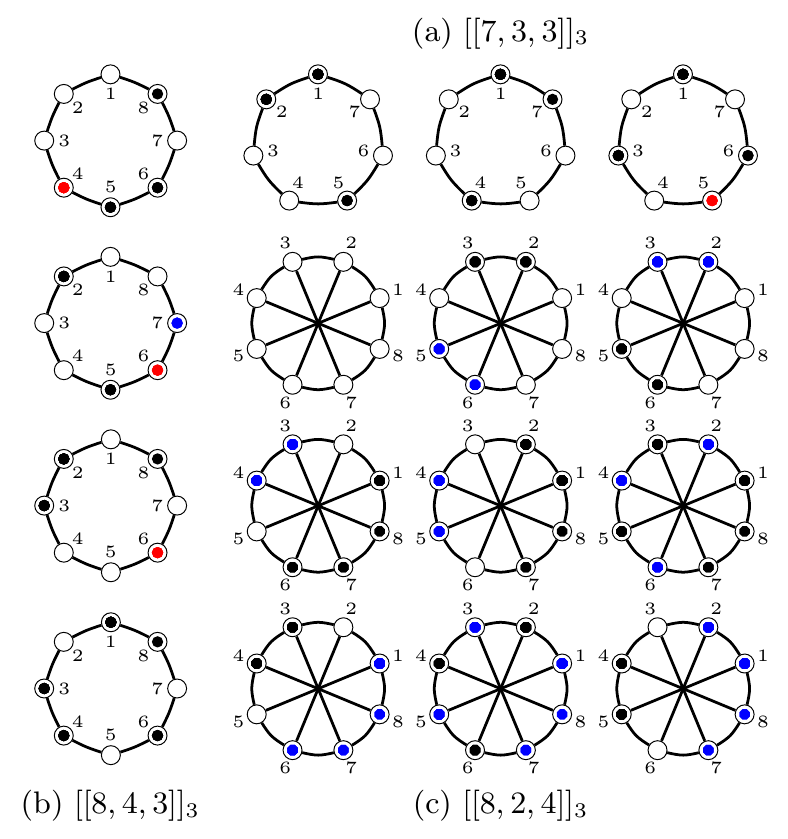} \caption{\label{tu3} (a) Three
generators of the coding group for the code $[[7,3,3]]_3$; (b) Four
generators of the coding group for the code $[[8,4,3]]_3$; (c) All
members of the coding groups for the code $[[8,2,4]]_3$.}
\end{figure}

We found also two families of optimal qutrit codes, namely the codes
$[[7,3,3]]_3$ and $[[8,4,3]]_3$, which can be constructed from the loop graphs
$L_7$ and $L_8$ respectively. The generators of the coding groups are indicated
in Fig.\ref{tu3}(a) and Fig.\ref{tu3}(b). For $d=3$ the blue and red vertices coincide.
However those generators of the coding groups are also valid for the codes
$[[7,3,3]]_p$ and $[[8,4,3]]_p$ for all odd $p>3$ which will be discussed in
details in the next section.

\subsection{The code $[[8,2,4]]_3$}

The last example is the code $[[8,2,4]]_3$ which is found only
recently \cite{looi} from the hype cube graph. The graph considered
here is the equal-weighted wheel graph $W_8$ as shown in
Fig.\ref{tu3}(c) and the corresponding graph state is denoted as
$|W_8\rangle$. The coding group is generated by
\begin{equation}
(0,1,1,0,2,2,0,0)\ \textrm{and}\ (1,0,2,2,0,1,1,1)
\end{equation}
and the graphical code $(W_8,2,4)_3$ is spanned by the graph-state basis
\begin{equation}
\{\mZ_1^b\mZ_2^a\mZ_3^{a+2b}\mZ_4^{2b}\mZ_5^{2a}\mZ_6^{2a+b}\mZ_7^{b}\mZ_8^{b}|W_8\rangle|a,b\in\mathbb Z_3\}
\end{equation}
The stabilizer of the code has the following set of generators
\begin{equation}
\langle \mG_1\mG_4, \mG_2\mG_5,\mG_3\mG_6,\mG_4\mG_7,\mG_4\mG_8,\mG_1\mG_2^2\mG_3
\rangle.
\end{equation}

To conclude this section we list all the stabilizer codes that
saturate the Singleton bound, i.e., $ k+2d\leq n+2, $ in Table
\ref{table 2}.
\begin{table}
\caption{\label{table 2}The codes for $p=3$.}
\begin{tabular}
 {l |l l l} \hline\hline
     &d=2&d=3&d=4\\
\hline
 n=3&$[[3,1,2]]_3$\\
 n=4&$[[4,2,2]]_3$\\
 n=5&$[[5,3,2]]_3$&$[[5,1,3]]_3$\\
 n=6&$[[6,4,2]]_3$&$[[6,2,3]]_3$\\
 n=7&$ $&$[[7,3,3]]_3$&$ $\\
 n=8&$ $&$[[8,4,3]]_3$&$[[8,2,4]]_3$\\
 \hline\hline
\end{tabular}
\end{table}
It should be emphasized that the absence of some codes in Table \ref{table
2} does not imply that these codes do not exist. To find some of these codes are
beyond our computers' capacity and some of them will be constructed in the
following sections. Furthermore, we can find the codes with a length $n\geq6$ via a
random searching method, thus some of absent codes may
still be found by using our systematic algorithm developed
above.

\section{ graphical Nonbinary QECCs via Analytical Constructions}

Because the clique finding problem is intrinsically an NP-complete
problem, it is not plausible to rely on the numerical search for
codes with larger length or higher dimension. Here we shall provide
some analytical constructions of good codes for higher dimension,
which is based on the graphs and their coding cliques in lower
dimensions found via computer research. In practice, we start from a
graphical code found for qutrit and then generalize the subspace to
arbitrary dimension by adopting the same graph and similar coding
clique, and finally we prove that the subspace provides a code in
arbitrary dimension.

\subsection{The nonadditive code $((3,p-1,2))_p$ with even $p$} {\it
Description of the code---} Suppose that $p=2q$. We consider the
star graph $S_3$ labeled with $V=\{A,B,C\}$ as shown in
Fig.\ref{tu4}(a). The $p-1$ dimensional subspace spanned by the
basis
\begin{equation}\label{b1}
\mZ_A^{l}\mZ_C^{2l}|S_3\rangle,\ \mZ_A^{q+j}\mZ_C^{2j+1}|S_3\rangle,
\end{equation}
with $0\leq l\le q-1$ and $0\le j\le q-2$ is a nonadditive code $((3,p-1,2))_p$.

{\it Proof---} It is enough to demonstrate that a subset $\mathbb
C_2^{p-1}$ composed of $2q-2$ vectors of forms $(l, 0, 2l )$ and
$(q+j,0 , 2j+1)$ satisfies all three conditions of coding clique.
Since the 2-purity set is empty and $l$ can assume value 0,  we have
only to show that the condition iii of coding clique is satisfied.
Any single qupit error can only cover those vectors of forms
$(a,b,a)$, $(b,c,0)$, and $(0,c,b)$ with $a,b,c\in\mathbb Z_p$ being
arbitrary. Considering the range of the $l$ and $j$ it is
straightforward to show that all vectors in $\mathbb C_2^{p-1}$ and
pairwise difference cannot be covered by single qubit error. Thus we
have a code $((3,p-1,2))_p$. The stabilizer of the code turns out to
be generated by $\mG_B$. Therefore it is a nonadditive code.

Via a similar construction by Rains \cite{rains} we are able to
construct the code $((2n+3,p^{2n}(p-1),2))_p$ for even $p$. We
consider the graph on $2n+3$ vertices composed of the star graph
$S_3$ as shown in Fig.\ref{tu4}(a) and the graph $B_{2n}$ as shown
in Fig.\ref{tu4}(b) and denote the corresponding graph state as
$|S_3\rangle\otimes|B_{2n}\rangle$. Let us denote by $\{|v\rangle\}$
the basis in Eq.(\ref{b1}) for the $((3,p-1,2))_p$ code constructed
above then the code subspace is spanned by the basis
\begin{equation}
(\mZ_A\mZ_C)^{\sum_{i=1}^ns_{2i}}(\mZ_B)^{\sum_{i=1}^ns_{2i-1}}|v\rangle\otimes\mZ^{\bs}|B_{2n}\rangle
\end{equation}
with $\bs\in\mathbb Z_p^{2n}$ being arbitrary. We notice that the phase flips
acting on $|v\rangle$ is a single qupit error on qupit $B$.

\subsection{The code $[[2n,2n-2,2]]$} We consider the graph on $2n$
vertices with all edges weighted 1 as shown in Fig.\ref{tu4}(b) and
denote the corresponding graph state as $|B_{2n}\rangle$. The
subspace spanned by the graph-state basis
\begin{equation}
\mZ_1^{-\sum_{j=1}^{n-1} a_j}\mZ_2^{-\sum_{j=1}^{n-1} b_j}\prod_{j=1}^{n-1}\mZ_{2j+1}^{a_j}
\prod_{j=1}^{n-1}\mZ_{2(j+1)}^{a_{j+1}}|B_{2n}\rangle
\end{equation}
with $a_j,b_j\in\mathbb Z_p$ for $j=1,2,\ldots n-1$ is the code $[[2n,2n-2,2]]$ whose stabilizer
is generated by
\begin{equation}
\begin{array}{c}
\mX_1\mZ_2\mX_3\mZ_4\ldots\mX_{2n-1}\mZ_{2n},\cr
\mZ_1\mX_2\mZ_3\mX_4\ldots\mZ_{2n-1}\mX_{2n}.
\end{array}
\end{equation}
It is straightforward to see from the stabilizer that every single
qupit error can be detected, i.e., not commute with at least one of
two generators defined above. In comparison in \cite{looi} the star
graph has been used to construct the code $[[n,n-2,2]]_p$.

\subsection{The code $[[2n+3,2n+1,2]]_p$ with odd $p$} 
\begin{figure}
\includegraphics{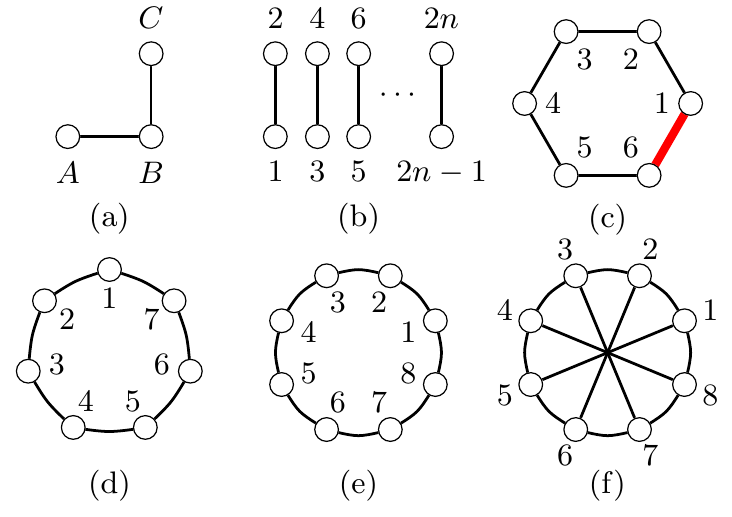}
\caption{\label{tu4}(a) The star graph $S_3$ on 3 vertices; (b) The
graph $B_{2n}$; (c) The loop graph $L_6$ on 6 vertices with 1 edge
weighted -1; (d) The loop graph $L_7$; (e) The loop graph $L_8$; (f)
The wheel graph $W_8$ on 8 vertices. }
\end{figure}

Consider the
graph on $2n+3$ composed of a subgraph $S_3$ as shown in
Fig.\ref{tu4}(a) and a subgraph $B_{2n}$ as shown in
Fig.\ref{tu4}(b). On the first 3 qupits a $[[3,1,2]]$ code can be
constructed for odd $p$ with the stabilizer given by $\langle
\mG_B,\mG_A\mG_C\rangle$ as shown previously. Then we obtain the
stabilizer of the code $[[2n+3,2n+1,2]]_p$ as
\begin{equation}
\begin{array}{c}
\mZ_A\mX_B\mZ_C\mX_1\mZ_2\mX_3\mZ_4\ldots\mX_{2n-1}\mZ_{2n},\cr
\mX_A\mZ_B^2\mX_C\mZ_1\mX_2\mZ_3\mX_4\ldots\mZ_{2n-1}\mX_{2n}.
\end{array}
\end{equation}

\subsection{The nonadditive code $((5,p,3))_p$ with $p>3$}

We consider the loop graph $L_5$ with all edges weighted 1 and the
subset $\mathbb C_3^p\subset\mathbb Z^5_p$ composed of $p$ vectors
including $(j,j,j,j,j)$ with $j\neq 2,-1$ and two additional vectors
$\ba=(2,-1,-1,2,-1)\ \textrm{and}\ \bb=(-1,2,2,-1,2).$ All of these
vectors have full support and obviously cannot be covered by any
single qupit error which is able to cover at most 3 vertices. All
vectors $(c_1,c_2,\cdots,c_5)$ with full support that can be covered
by a 2-qupit error has the property
\begin{equation}\label{c5}
c_n+c_{n+1}=c_{n+3}.
\end{equation}
It is obvious that all vectors in
$\mathbb C_3^p$ do not have such a property. Furthermore the
difference of any pair of vectors in $\mathbb C_3^p$ has full
support and do not have the property Eq.(\ref{c5}) either. Therefore
$\mathbb C_3^p$ is a coding clique since conditions i and ii are
trivially satisfied. To find all the stabilizer of the code
constructed from the coding clique $\mathbb C_3^p$, i.e., subspace
spanned by the graph-state basis
\begin{equation}\label{code5}
\{(\mZ_1\mZ_2\mZ_3\mZ_4\mZ_5)^j|L_5\rangle,\ \mZ^\ba|L_5\rangle,\
\mZ^\bb|L_5\rangle\},
\end{equation}
we have to solve equations
in Eq.(\ref{cs}). It turns out that the
total number of the solutions to Eq.(\ref{cs}) is $q^{3}$ if $p$ is not a multiple of 3
and $3q^3$ if $p$ is not 3 but can be divided by 3. In either case the number of
stabilizers is less than $p^4$ if $p>3$ and since the code saturates the Singleton
bound we conclude that the code Eq.(\ref{code5}) is nonadditive.

\subsection{The  code $[[6,2,3]]_p$}

{\it Description of the code---} Consider the weighted loop graph on
6 vertices as shown in Fig.\ref{tu4}(c) and the corresponding graph state
$|L_6\rangle$. The $p^2$-dimensional subspace spanned by the
graph-state basis
\begin{equation}
\left\{\mZ_1^a\mZ_2^{a+b}\mZ_3^{b}\mZ_4^{-a}\mZ_5^{a-b}\mZ_6^b|L_6\rangle\big|
a,b\in\mathbb Z_p\right\}
\end{equation}
is a $[[6,2,3]]_p$ code for all odd $p>2$. A set of the generators
of its stabilizer is listed in Table \ref{t6}.
\begin{table}
\caption{\label{t6}The stabilizer of the code $[[6,2,3]]_p$}
\begin{equation*}
\begin{array}{r|cccccc}
\hline\hline
&1&2&3&4&5&6\cr\hline
\mG_1\mG_4          &\mX&\mZ&\mZ&\mX&\mZ&\bar\mZ\cr
\mG_3\bar\mG_6&\mZ&\mZ&\mX&\mZ&\bar\mZ&\bar\mX\cr
\mG_1\bar\mG_2\mG_3&\bar\mZ\mX&\bar\mX\mZ^2&\bar\mZ\mX&\mZ&\mI&\bar\mZ\cr
\mG_3\mG_4\mG_5&\mI&\mZ&\mZ\mX&\mX\mZ^2&\mZ\mX&\mZ\cr
\hline\hline
\end{array}
\end{equation*}

\end{table}

{\it Proof---} Because of Theorem 1 we have only to prove that the
subset of vectors
\begin{equation}
\{\bc=(a,a+b,b,-a,a-b,b)|a,b\in\mathbb Z_p\}
\end{equation}
satisfies all three conditions of the coding clique. Conditions i)
and ii) are obviously satisfied since $a,b$ can assume value zero and the
3-purity set is empty for the loop graph for any dimension $p$. To prove condition iii)
we have to show that $\bc$ cannot be covered by single and 2-qupit errors for all
$a,b\in\mathbb Z_p$. Obviously $\bc$ cannot be covered by any single
vertex error, we consider only 2-qupit errors in what follows.
\begin{itemize}
\item[i)]$b=0, a\neq0$ with $\bc=(a,a,0,-a,a,0)$. If a 2-qupit error is supported on vertices
$\{1,2,4,5\}$ then it can only takes form
$$
\mX_{1}^{n}\mZ_{1}^{s}\mX_{5}^n\mZ_{5}^{t},\quad
\mX_{2}^{n}\mZ_{2}^{s}\bar\mX_{4}^{n}\mZ_{4}^{t},\quad
\mX_{3}^{s}\mX_{6}^{t},
$$
which covers $(s,n,0,n,t,0)$, $(n,s,0,t,-n,0)$ and $(-s,t,0,t,s,0)$
respectively with $n, s, t \in \mathbb Z_p$ being arbitrary. All these 3 types of vectors
cannot be identified with $\bc$ provided $p>2$ and odd. For example if
$(s,n,0,n,t,0)=\bc$ then $2a=0$ which is impossible for odd $p$.

\item[ii)] $a=0, b\neq0$ with $\bc=(0,b,b,0,-b,b)$. In this case
it can be proved in exactly the same
manner as the first case that $\bc$ cannot be covered by 2-qupit
error.

\item[iii)]$a=b\ne 0$ with $\bc=(a,2a,a,-a,0,a)$.
Any 2-qupit error that covers $\{1,2,3,4,6\}$ can only take form
$\mX_1^m\mZ_1^s\mX_3^n\mZ_3^t$ or $\mX_2^m\mZ_2^s\mX_5^n,$
which covers $$(s,m+n,t,n,0,-m)\ \textrm{or}\ (m,s,m,n,0,n)$$
respectively with
$m,n,s,t\in \mathbb Z_p$ being arbitrary. To cover
$\bc$ we have to require $2a=-2a$ or $a=-a$ which are impossible for
odd $p$.

\item[iv)]$a=-b\ne0$ with $\bc=(-b,0,b,b,-2b,b)$.
Any 2-qupit error that covers 5 vertices $\{1,3,4,5,6\}$ can only be
$\mX_2^m\mX_5^n\mZ_5^s$ or $\mX_4^m\mZ_4^s\mX_6^n\mZ_6^t,$ which
covers $$(m,0,m,n,s,n)\ \textrm{or}\ (-n,0,m,s,m+n,t)$$ respectively with
$m,n,s,t\in \mathbb Z_p$ being arbitrary. To cover
$\bc$ we have to require $b=-b$ or $2b=-2b$ which is impossible for odd
p.

\item[v)] $a\neq\pm b, a,b\ne 0$ with $\bc=(a,a+b,b,-a,a-b,b)$.\\
To cover all the 6 vertices a 2-qupit error can only takes form
$$\mX_1^m\mZ_1^s\mX_4^n\mZ_4^t,\quad \mX_2^m\mZ_2^s\mX_5^n\mZ_5^t,\quad
\mX_3^m\mZ_3^s\mX_6^n\mZ_6^t,$$ which covers $(s,m,n,t,n,-m)$,
 $(m,s,m,n,t,n)$, and $(-n,m,s,m,n,t)$ respectively. However all these vectors
 can not be identified with $\bc$ if $p$ is odd. For example if $(s,m,n,t,n,-m)=\bc$
 then $2a=0$ which is impossible for odd p.
\end{itemize}
In summary, for any $a,b\in\mathbb Z_p$ the vector $\bc$ cannot be
covered by any single or 2-qupit errors so that is a coding clique
and corresponding subspace is the $[[6,2,3]]_p$ code for all odd
$p>2$.

\subsection{The  code $[[7,3,3]]_p$}

{\it Description of the code---} Consider the equal weighted loop
graph on 7 vertices as shown in Fig.\ref{tu4}(d) and the corresponding graph
state $|L_7\rangle$. The $p^3$-dimensional subspace spanned by
the graph-state basis
\begin{equation}
\left\{\mZ_1^{a+b+c}\mZ_2^{a}\mZ_3^{c}\mZ_4^{b}\mZ_5^{a-c}\mZ_6^{-c}\mZ_7^b|L_7\rangle\big|
a,b,c\in\mathbb Z_p\right\}
\end{equation}
is a $[[7,3,3]]_p$ code for all odd $p>2$. A set of the generators
of its stabilizer is listed in Table \ref{t7}.
\begin{table}[tbph]
\caption{\label{t7}The stabilizer of the code $[[7,3,3]]_p$}
\begin{equation*}
\begin{array}{r|ccccccc}
\hline\hline
&1&2&3&4&5&6&7\cr\hline
\mG_3\mG_6          &\mI&\mZ&\mX&\mZ&\mZ&\mX&\mZ\cr
\bar\mG_4\mG_7&\mZ&\mI&\bar\mZ&\bar\mX&\bar\mZ&\mZ&\mX\cr
\bar\mG_2\mG_3\mG_5&\bar\mZ&\bar\mX\mZ&\bar\mZ\mX&\mZ^2&\mX&\mZ&\mI\cr
\bar\mG_1\mG_2\mG_3\mG_4&\bar\mX\mZ&\mX&\mX\mZ^2&\mZ\mX&\mZ&\mI&\bar\mZ\cr
\hline\hline
\end{array}
\end{equation*}
\end{table}

{\it Proof---} Because of Theorem 1,
we have to show that the subset of $\mathbb Z_p^7$ defined as
\begin{equation}\label{7c}
\{\bc=(a+b+c,a,c,b,a-c,-c,b)|a, b, c \in\mathbb{Z}_p\}
\end{equation}
is a coding clique. Since the 3-purity set is empty for the loop
graphs with any dimension $p$, we need only to show that any nonzero
$\bc$ cannot be covered by any single or 2-qupit error.

It is not difficult to see that $\bc$ cannot be covered by any
single qupit error, which covers a vectors with 4 consecutive
components being zero. Because of the symmetry of the loop graph,
all 2-qupit errors can be classified into 3 types $\mathbb
T_{12},\mathbb T_{13}$, and $\mathbb T_{14}$ where $\mathbb T_{lj}$
denotes two errors occur on qupits $n+l$ and $n+j$ with $n\in\mathbb
Z_7$. Each type of errors covers the vector $(c_1,c_2,\ldots,c_7)$
with properties
\begin{itemize}

\item[$\mathbb T_{12}$:] $c_{n}=c_{n+1}=c_{n+2}=0$;

\item[$\mathbb T_{13}$:] $c_{n-1}+c_{n+3}=c_{n+1}$ and
$c_{n-2}=c_{n-3}=0$;

\item[$\mathbb T_{14}$:] $c_{n-1}=c_{n+1}$, $c_{n-2}=0$,
and $c_{n-3}=c_{n+2}$.

\end{itemize}
It can be easily checked that as given in Eq.(\ref{7c}) for
arbitrary $a, b, c \in\mathbb{Z}_p$ does not belong to all these 3
types of vectors when $p$ is odd. For example a $\mathbb T_{12}$
type of error covers  vector with 3 consecutive  components being
zero, which is impossible to be identified with a nonzero $\bc$.
That is to say $\bc$ cannot be covered by any 2-qupit errors and
therefore we have a coding clique and a $[[7,3,3]]_p$ code for all
odd $p$.

\subsection{The code $[[8,4,3]]_p$}

{\it Description of the code---} Consider the equal weighted loop
graph on 8 vertices as shown in Fig.\ref{tu4}(e) and the corresponding graph
state $|L_8\rangle$. The $p^4$-dimensional subspace spanned by
the graph-state bases
\begin{equation}
\mZ_1^{e}\mZ_2^{b-c}\mZ_3^{e-c}\mZ_4^{e-a}\mZ_5^{a+b}\mZ_6^{a-b+c+e}\mZ_7^{2b}\mZ_8^{a-c+e}|L_8\rangle
\end{equation}
with $a,b,c,e\in\mathbb Z_p$ is a $[[8,4,3]]_p$ code for all odd $p>2$.
A set of the generators of its stabilizer is listed in Table~\ref{t83}.
\begin{table}[tbph]
\caption{\label{t83}The stabilizer of the code $[[8,4,3]]_p$}
\begin{equation*}
\begin{array}{r|cccccccc}
\hline\hline
&1&2&3&4&5&6&7&8\cr\hline
\bar\mG_1\bar\mG_3\mG_4\mG_8      &\bar\mX\mZ&\mI&\mX\mZ&\mZ\mX&\mZ&\mI&\mZ&\bar\mZ\mX\cr
\mG_2^{2}\bar\mG_3\bar\mG_5\mG_6  &\mZ^2&\mX^2\bar\mZ&\mZ^2\bar\mX&\bar\mZ^{2}&\bar\mX\mZ&\bar\mZ\mX&\mZ&\mI\cr
\bar\mG_1^{2}\mG_4^2\mG_5^2\bar\mG_7&\bar\mX^{2}&\bar\mZ^{2}&\mZ^2&\mX^2\mZ^2&\mZ^2\mX^2&\mZ&\mX&\bar\mZ^3\cr
\bar\mG_1^{2}\bar\mG_2\mG_3\mG_4\mG_5 &\bar\mZ\bar\mX^{2} &\bar\mZ\bar\mX &\mX &\mX\mZ^2 &\mX\mZ&\mZ&\mI&\bar\mZ^2\cr
\hline\hline
\end{array}
\end{equation*}
\end{table}

{\it Proof---}Because of Theorem 1,
we have to show that the subset of $\mathbb Z_p^8$ composed of vectors of form
\begin{equation}\label{c843}
\bc=(e, b-c, e-c, e-a, a+b, a-b+c+e, 2b, a-c+e)
\end{equation} which
with $a,b,c,e\in\mathbb Z_p$ is a coding clique of the loop graph.
Because of the symmetry of the loop graph, all 2-qupit errors can be
classified into 4 types $\mathbb T_{12},\mathbb T_{13}$, $\mathbb
T_{14}$, and $\mathbb T_{15}$. Each type of errors covers the vector
$(c_1,c_2,\ldots,c_8)$ with properties ($n\in\mathbb Z_8$)
\begin{itemize}
\item[$\mathbb T_{12}$:] $c_{n+1}=c_{n+2}=c_{n+3}=c_{n+4}=0$;

\item[$\mathbb T_{13}$:] $c_{n+2}=c_{n+3}=c_{n+4}=0$ and $
c_{n+1}+c_{n-3}=c_{n-1}$;

\item[$\mathbb T_{14}$:]
$c_{n+2}=c_{n+3}=0$, $c_{n+1}=c_{n-1}$, and $c_{n-2}=c_{n-4}$;

\item[$\mathbb T_{15}$:] $c_{n}=c_{n+4}=0$, $c_{n+1}=c_{n+3}$, and $c_{n-1}=c_{n-3}$.

\end{itemize}
As an example we consider a $\mathbb T_{12}$ type of errors which
covers a vector with 4 consecutive components being zero. A nonzero
vector $\bc$ as defined in Eq.(\ref{c843}) is not possible to have
such a property, e.g., the last 4 components being zero. In the same
manner it can be checked that $\bc$ does not belong to all those 4
types of vectors above for arbitrary $a,b,c,e\in\mathbb Z_p$ when
$p$ is odd. Since all single qupit errors cover vectors with the
same property as in type $\mathbb T_{12}$, we conclude that $\bc$
cannot be covered by any single or 2-qupit errors so that we have a
coding clique and therefore a $[[8,4,3]]_p$ code for all odd $p$.

\subsection{The code $[[8,2,4]]_p$}

{\it Description of the code---} Consider the equal weighted wheel graph
on 8 vertices as shown in Fig.3(d) and the corresponding graph state
$|W_8\rangle$. The $p^2$-dimensional subspace spanned by the
graph-state basis
\begin{equation}
\left\{\mZ_1^{b}\mZ_2^{a}\mZ_3^{a-b}\mZ_4^{2b}\mZ_5^{2a}\mZ_6^{b-a}
\mZ_7^b\mZ_8^b|\Gamma\rangle\big| a,b\in\mathbb Z_p\right\}
\end{equation}
is a $[[8,2,4]]_p$ code for all odd $p>2$. A set of the generators
of its stabilizer is listed in Table~\ref{t84}.
\begin{table}[tbph]
\caption{\label{t84}The stabilizer of the code $[[8,2,4]]_p$}
\begin{equation*}
\begin{array}{r|cccccccc}
\hline\hline
&1&2&3&4&5&6&7&8\cr\hline
\mG_1\bar\mG_7      &\mX&\mZ&\bar\mZ&\mI&\mZ&\bar\mZ&\mX&\mI\cr
\mG_1\bar\mG_8      &\mX\bar\mZ&\mZ&\mI&\bar\mZ&\mZ&\mI&\bar\mZ&\mZ\bar\mX\cr
\mG_1^{2}\bar\mG_4  &\mX^2&\mZ^2&\bar\mZ&\bar\mX&\mZ&\mI&\mI&\mZ\cr
\mG_2^2\bar\mG_5    &\mZ&\mX^2&\mZ^2&\bar\mZ&\bar\mX&\mZ&\mI&\mI\cr
\mG_3\mG_6          &\mI&\mZ^2&\mX &\mZ&\mZ&\mX&\mZ^2&\mI\cr
\mG_1\bar\mG_2\mG_3 &\bar\mZ\mX &\bar\mX\mZ^2 &\bar\mZ\mX &\mZ &\mZ&\bar\mZ&\mZ&\mZ\cr
\hline\hline
\end{array}
\end{equation*}
\end{table}

{\it Proof---} Because of Theorem 1 we have to show that the subset of $\mathbb Z_p^8$
defined as
\begin{equation}\label{c84}
\{\bc=(b,a,a-b,2b,2a,b-a,b,b)|a, b \in\mathbb Z_p\}.
\end{equation}
is a coding clique of the wheel graph. We have only to prove
condition iii of the coding clique since the 4-purity set is empty.

A single qupit  error on vertex $n$ can cover a vector with property
$c_{n\pm 2}=c_{n\pm 3}=0$ which is impossible for $\bc$ defined
above. Because of symmetry all 2-qupit errors can be classified into
4 types $\mathbb T_{12},\mathbb T_{13}$, $\mathbb T_{14}$, and
$\mathbb T_{15}$ with each type of errors covering the vector
$(c_1,c_2,\ldots,c_8)$ with properties ($n\in\mathbb Z_8$)
\begin{itemize}
\item[$\mathbb T_{12}$:]$c_{n+1}=c_{n+4}$, $c_{n+2}=c_{n+5}=0$, and
$c_{n+3}=c_{n+6}$;

\item[$\mathbb T_{13}$:]$c_{n}=c_{n+3}$  and $c_{n+1}=c_{n+4}$;

\item[$\mathbb T_{14}$:]$c_{n}=c_{n+3}$  and $c_{n+1}=c_{n+2}=0$;

\item[$\mathbb T_{15}$:]$c_{n}=c_{n+6}$, $c_{n+2}=c_{n+4}$ and $c_{n+1}=c_{n+5}=0$.
\end{itemize}
It can be checked in a straightforward manner that none of the above
equalities can be satisfied by $\bc$ ad defined in Eq.(\ref{c84}).
For example we consider error of type $\mathbb T_{12}$ with $n=0$.
In this case we have constraints $b=2b$, $a=2a$, and $a-b=b-a$ which
are impossible for nonzero $\bc$ when $p$ is odd. This is true for
all $n\in\mathbb Z_8$. Thus any 2-qupit error cannot cover $\bc$.

All 3-qupit errors can be classified into 7 types $\mathbb
T_{123},\mathbb T_{124}$, $\mathbb T_{125}$, $\mathbb T_{126}$,
$\mathbb T_{127}$, $\mathbb T_{135}$, and  $\mathbb T_{136}$ with
each type of errors covering vectors with the following properties
($n\in\mathbb Z_8$)
\begin{itemize}
\item[$\mathbb T_{123}$:]$c_{n+2}=c_{n+5}$  and $c_{n-2}=c_{n-5}$;

\item[$\mathbb T_{124}$:]$c_{n}=c_{n+3}$  and $c_{n+2}=0$;

\item[$\mathbb T_{125}$:]$c_{n}+c_{n+1}=c_{n+3}$  and $c_{n+4}=0$;

\item[$\mathbb T_{126}$:]$c_{n}+c_{n-1}=c_{n-3}$  and $c_{n-4}=0$;

\item[$\mathbb T_{127}$:]$c_{n}=c_{n-3}$  and $c_{n-2}=0$;

\item[$\mathbb T_{135}$:]$c_{n}+c_{n+1}=c_{n+3}$  and
$c_{n}+c_{n-1}=c_{n-3}$;

\item[$\mathbb T_{136}$:] $c_{n}=c_{n+2}+c_{n+3}=c_{n-2}+c_{n-3}$.
\end{itemize}
It can be checked in a tedious but straightforward manner that for
every $n\in\mathbb Z_8$ none of those equalities above can be
satisfied by $\bc$ as defined in Eq.(\ref{c84}). For an example we
consider the 3-qupit error of type $\mathbb T_{126}$ with $n=1$. In
this case we have $b+b=b-a$ and $2a=0$ which are impossible for
nonzero $\bc$.

To summarize, the vector $\bc$ defined in Eq.(\ref{c84}) for all
$a,b\in\mathbb Z_p$ is $4$-uncoverable. Thus we have proved that the
subspace defined in is a $[[8,4,2]]_p$ code for all odd dimension
$p$.

\section{Codes from composite systems }

Consider a system with $pq$ levels whose computational bases are
denoted by $\{|l\rangle_{pq}\}_{l=0}^{pq-1}$. We can also regard
this system as a composite system of a $p$-level system and a
$q$-level system, whose computational bases are denoted as
$\{|s\rangle_{p}\}_{s=0}^{p-1}$ and $\{|t\rangle_{q}\}_{t=0}^{q-1}$
respectively. If there are $n$ copies of $pq$-level systems, we also
have $n$ copies of $p$-level and $q$-level systems. On the other
hand if we have two groups of $p$-level and $q$-level systems we can
also obtain $n$ copies of $pq$-level system by pairing up one
$p$-level system and one $q$-level system to made up a composite
system.

Given a $\mathbb Z_p$-weighted graph $(V,\Gamma_p)$ and a $\mathbb
Z_q$-weighted graph $(V,\Gamma_q)$ on the same vertex set $V$ and
corresponding coding cliques $\mathbb C_d^K$ and  $\tilde{\mathbb
C}_d^{\tilde K}$, the subspace spanned by the basis
\begin{equation}
\left\{\mZ^{\bc}|\Gamma_p\rangle\otimes\mZ^{\tilde\bc}|\Gamma_q\rangle\Big|\bc\in
\mathbb C_d^K,\tilde\bc\in \tilde{\mathbb C}_d^{\tilde K}\right\}
\end{equation}
is an $((n,K\tilde K,d))_{pq}$ code with $n=|V|$. This is because
all the direct products of Pauli errors of the $p$-level systems and
$q$-level systems form a nice error basis for the $pq$-level system.
That is to say, via a direct product of two graphical codes
$(G,K,d)_{p}$ and $(\tilde G,\tilde K,d)_{q}$ we can construct a
code $((n,K\tilde K,d))_{pq}$, which is however not necessarily to
be another graphical code. As will be shown below this construction
will yield a graphical code of a higher dimension if $p$ and $q$ are
coprime, in which case there exist two integers $\alpha$ and $\beta$
such that
 \begin{equation}
 \alpha p+\beta q=1.
 \end{equation}

Given a $\mathbb Z_{pq}$-weighted graph $(V,\Gamma_{pq})$ we can
build  a $\mathbb Z_p$-weighted graph $(V,\Gamma_p)$ and a $\mathbb
Z_q$-weighted graph $(V,\Gamma_q)$ whose adjacency matrices are
given by
\begin{equation}\label{gama1}
p\Gamma_{pq}\equiv\Gamma_q\ (\textrm{mod}\ q),\quad q\Gamma_{pq}\equiv\Gamma_p
\ (\textrm{mod}\ p).
\end{equation}
On the other hand, given a $\mathbb Z_p$-weighted graph
$(V,\Gamma_p)$ and a $\mathbb Z_q$-weighted graph $(V,\Gamma_q)$ on
the same vertex set $V$, we can also build a $\mathbb
Z_{pq}$-weighted graph $(V,\Gamma_{pq})$ with adjacency matrix given
by
 \begin{equation}\label{gama2}
 \Gamma_{pq}\equiv p\,\alpha^2\Gamma_q+q\, \beta^2\Gamma_p \ (\mbox{mod}\ pq).
 \end{equation}
By relabeling of the bases of a $pq$-level system according to
$|s\rangle_p\otimes|t\rangle_q\mapsto |pt+qs\rangle_{pq}$, we can
define an isometry between $n$ $pq$-level systems and $n$ pairs of
$p$-level subsystems and $q$-level subsystems as
\begin{equation}\label{r}
\mR=\sum_{\bs\in\mathbb Z_p^V}\sum_{\bt\in\mathbb
Z_q^V}|p\bt+q\bs\rangle_{pq}\langle \bs|_p\otimes\langle \bt|_q,
\end{equation}
which is only possible when $p$ and $q$ are coprime, we have
\begin{equation}
|\Gamma_{pq}\rangle=\mR|\Gamma_{q}\rangle.
\end{equation}
Accordingly the bit flips and phase flips are related to each other
via
\begin{equation}
\mX_{pq}=\mR(\mX_p^\beta\otimes\mX_q^\alpha)\mR^\dagger,\quad
\mZ_{pq}=\mR(\mZ_p\otimes\mZ_q)\mR^\dagger.
\end{equation}

\textbf{Theorem 2} If $p,q$ are coprime then the graph state on
$\mathbb Z_{pq}$-weighted graph is in a one-to-one correspondence
with the direct product of two graph states on a $\mathbb
Z_p$-weighted graph and a $\mathbb Z_q$-weighted graph  whose
adjacency matrices are related via Eqs.(\ref{gama1},\ref{gama2}).

According to the fundament theorem of arithmetics any integer $p$ can be expressed
as
$p=p_1^{n_1}p_2^{n_2}\ldots
p_L^{n_L}$
for some primes $p_i$ $(i=1,\ldots,L)$. Therefore a graph state on the
subsystems of dimension $p$ will be in a one-to-one correspondence
of a direct product of graph states on the subsystems of dimension
$p_i^{n_i}$. As a consequence we have only to consider the graph states of
systems of a dimension that is prime or a power of prime. So far the definition
of graph state is unique in the case of
prime dimension and there is a second definition of graph states
\cite{grsl} in the case of prime power dimension.

\begin{figure}
\includegraphics{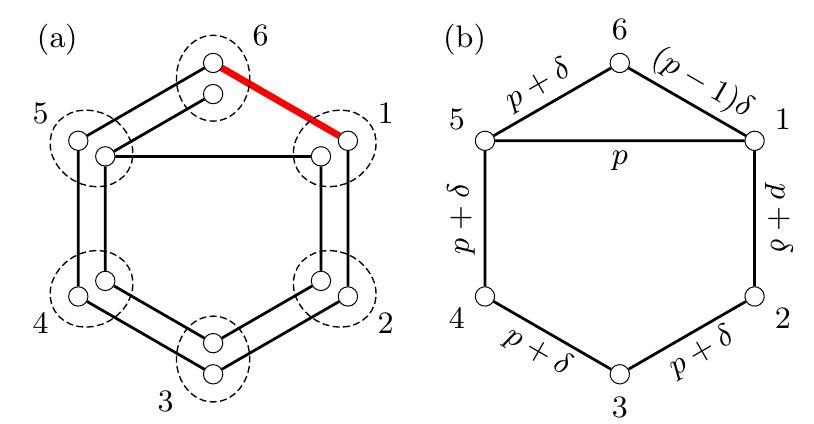} \caption{An example of the composite system codes.
(a) The inner graph is a $\mathbb Z_{2}$-weighted graph and the
outer graph is a $\mathbb Z_{p}$-weighted graph. (b) The $\mathbb
Z_{2p}$-weighted graph $H_6$ where $\delta=(p+1)^2/2$.}
\end{figure}

As an example we consider the case where $q=2$ and there 6 copies of
$2p$-level composite systems and accordingly 6 pairs of qubits and
qupits. A direct product of a graphical code $[[6,1,3]]_2$ on qubits
constructed from the loop graph with a leaf \cite{SiXia} as shown in
Fig.5(a) and the graphical code $[[6,2,3]]_p$ arising from the graph
$L_6$ as constructed in Sec.V(E) will result a code $((6,2\cdot
p^2,3))$. It is in fact a stabilizer code and we should note that
the dimension of the coding subspace is not a power of $2p$.

Via the relabeling of the bases of the $2p$-level systems as
specified $\mR$ in Eq.(\ref{r}) we obtain a $\mathbb
Z_{2p}$-weighted graph $H_6$ as shown in Fig.5(b). From this graph
we consider a subset $\mathbb C$ of $2p^2$ vectors in $\mathbb
Z_{2p}$ of form
\begin{equation}
(p+1)(a,a+b,b,-a,a-b,b)+p(c,c,c,c,c,0)
\end{equation}
with $a,b\in \mathbb Z_p$ and $c\in \mathbb Z_2$. The subspace
spanned by the graph-state basis $\{\mZ^\bc|H_6\rangle|\bc\in\mathbb
C\}$ is exactly the stabilizer code $((6,2p^2,3))$ constructed from
composite systems. A set of  the generators of its stabilizer is
listed in Table VI.
\begin{table}[tbph]
\caption{\label{t62}The stabilizer of the code $((6,2p^2,3))_{2p}$.}
\begin{equation*}
\begin{array}{r|cccccc}
\hline\hline &1&2&3&4&5&6\cr\hline
\mG_1^2\mG_4^2 &\mX^2&\mZ^{p+1}&\mZ^{p+1}&\mX^{2}&\mZ^{p+1}&\bar\mZ^{p+1}\cr
\mG_3^2\bar\mG_6^2&\mZ^{p+1}&\mZ^{p+1}&\mX^2&\mZ^{p+1}&\bar\mZ^{p+1}&\bar\mX^2\cr
\mG_1^2\bar\mG_2^2\mG_3^2&\bar\mZ^{p+1}\mX^2&\bar\mX^2\mZ^2&\bar\mZ^{p+1}\mX^2&\mZ^{p+1}&\mI&\bar\mZ^{p+1}\cr
\mG_3^2\mG_4^2\mG_5^2&\mI&\mZ^{p+1}&\mZ^{p+1}\mX^2&\mX^2\mZ^{2}&\mZ^{p+1}\mX^2&\mZ^{p+1}\cr\hline
\mG_6^p &\mI&\mI&\mI&\mI&\mZ^p&\mX^p\cr
\mG_1^p\mG_2^p &\mX^p\mZ^p&\mZ^p\mX^p&\mZ^p&\mI&\mZ^{p}&\mI\cr
\mG_1^p\mG_3^p &\mX^p&\mI&\mX^p&\mZ^p&\mZ^{p}&\mI\cr
\mG_1^p\mG_4^p &\mX^p&\mZ^p&\mZ^{p}&\mX^p&\mI&\mI\cr
\mG_1^p\mG_5^p &\mX^p\mZ^p&\mZ^p&\mI&\mZ^{p}&\mZ^p\mX^p&\mZ^{p}\cr
\hline\hline
\end{array}
\end{equation*}
\end{table}

Because the code $[[6,2,3]]_2$ does not exist we cannot remove any
stabilizers from the lower half part of the above table to obtain a
$(2p)^2$-dimensional code space, therefore the code
$((6,2p^2,3))_{2p}$ is not a subcode of a $[[6,2,3]]_{2p}$, if there
is any.

\section{Discussion}

We have generalized the graphical construction \cite{SiXia} of QECCs
to the nonbinary case to find both additive and nonadditive codes
based on nonbinary graph states.  The advantages of our graphical
approach lies in the fact that we are able to construct codes on
physical systems of arbitrary dimension, prime or nonprime, to
construct both additive or nonadditive codes, pure or impure. In
addition the basis for all codes are explicitly constructed. In
principle for prime dimension our method exhausts all the stabilizer
codes, which can be demonstrated in exactly the same manner as in
binary case \cite{SiXia}.

Via numerical search we have found many optimal codes including two
optimal codes $[[8,4,3]]_3$ and $[[8,2,4]]_3$ which have been found
in \cite{looi} on a hyper cube graph. Since the clique finding
problem for non-binary case is even more difficult and is
intrinsically a NP-complete problem, our computational result is
mainly limited in codes with $n\leq8$ and small distance $d\leq4$.
With the help of the codes found numerically we also manage to
construct analytically some families of optimal stabilizer codes
that saturating the Singleton bound such as $[[6,2,3]]_p$,
$[[7,3,3]]_p$, $[[8,4,3]]_p$ and $[[8,2,4]]_p$ for any odd $p$. We
have also explicitly constructed the code $[[n,n-2,2]]_p$ except the
case of even $p$ and odd $n$. There exist stabilizer codes whose
code subspace is not a power of the dimension of physical systems
such as the code $((6,2\cdot p^2,3))_{2p}$ which is constructed via
a composite system approach. Furthermore we have constructed a
family of nonadditive codes $((3,p-1,2))_p$ with even $p$, a
nonadditive optimal code $((5,p,3))_p$ for all $p>3$.

Finally, we briefly address some questions which are still open.
Although our method can also be used to find additive and
nonadditive codes for non-prime dimension, we can not exhaust all
the non-prime stabilizer codes because of our definition of graph
states. Because of Theorem 2 we have to consider the graph states on
system of prime and prime power dimension. There are at least two
different definitions of the graph state on $n$ subsystems of a
dimension $p=q^m$ being a power of prime: one is as defined in this
paper and in \cite{looi} and the other one is as defined  in
\cite{grsl}. A different definition of graph states may result in
some new families of codes.

\section*{Acknowledgement}

The financial supports from NNSF of China (Grant No. 10675107 and
Grant No. 10705025) and WBS (Project Account No): R-144-000-189-305,
Quantum information and Storage (QIS) are acknowledged.

\end{document}